\documentclass[11pt,pre,epsfig,address,onecolumn]{revtex4}%
\usepackage[english]{babel}
\usepackage[latin1]{inputenc}
\usepackage[dvips]{graphicx}
\usepackage{amsmath}
\usepackage{amssymb}
\usepackage{epsfig}
\usepackage{array}
\usepackage{amsfonts}
\usepackage{graphicx}
\usepackage{color}
\usepackage{subfigure}%
\newcommand{\be}{\begin{equation}}
\newcommand{\ee}{\end{equation}}
\newcommand{\bel}[1]{\begin{equation}\label{#1}}
\newcommand{\bea}{\begin{eqnarray}}
\newcommand{\eea}{\end{eqnarray}}
\newcommand{\balign}{\begin{align}}
\newcommand{\ealign}{\end{align}}
\newcommand{\ba}{\begin{array}}
\newcommand{\ea}{\end{array}}
\newcommand{\bfig}{\begin{figure}}
\newcommand{\efig}{\end{figure}}

\begin{document}
\title{Unusual shock wave in two-species driven systems with an umbilic point}
\author{Vladislav Popkov$^{1,2}$ and Gunter M. Sch\"utz$^{3,4}$ }
\affiliation{$^{1}$ Max Planck Institute for Complex Systems, N\"othnitzer Stra{ß}e 38,
01187 Dresden, Germany}
\affiliation{$^{2}$ Dipartimento di Fisica e Astronomia, Universit\`a di Firenze, via G.
Sansone 1, 50019 Sesto Fiorentino, Italy}
\affiliation{$^{3}$ Theoretical Soft Matter and Biophysics, Institute of Complex Systems
II, Forschungszentrum J\"ulich, 52425 J\"ulich, Germany}
\affiliation{$^{4}$Interdisziplin\"ares Zentrum f\"ur Komplexe Systeme, Universit\"at Bonn,
Br\"uhler Stra{ß}e 7, 53119 Bonn, Germany}
\date{\today}

\begin{abstract}
Using dynamical Monte-Carlo simulations we observe the occurrence of an
unexpected shock wave in driven diffusive systems with two conserved species
of particles. This U-shock is microscopically sharp, but does not satisfy the
usual criteria for the stability of shocks. Exact analysis of the large-scale
hydrodynamic equations of motion reveals the presence of an umbilical point
which we show to be responsible for this phenomenon. We prove that such an
umbilical point is a general feature of multi-species driven diffusive systems
with reflection symmetry of the bulk dynamics. We argue that an U-shock will
occur whenever there are strong repulsive interactions such that the
current-density relation develops a double-well and the umbilical point
becomes isolated.

\end{abstract}

\maketitle


\section{Introduction}

\label{sec::Introduction}

Nonequilibrium lattice gas models of interacting particles with noisy dynamics
\cite{Ligg99,Schu01} are paradigmatic models of systems far from equilibrium
and find a wide range of applications in biological, social and physical
contexts \cite{Muka00,Evan05,Scha10}. Driving forces due to bulk fields or
boundary gradients lead to steady state currents that invalidate the condition
of detailed balance and give rise to remarkable features which have no
equilibrium counterparts. As examples we mention boundary driven phase
transitions, spontaneous symmetry breaking and hysteresis in one spatial
dimension. Particles systems with two or more conserved species exhibit
particularly rich behaviour \cite{Schu03}.

The coarse grained space-time evolution of bulk-driven systems is governed by
two fundamental types of excitations: shocks, which carry discontinuities, and
rarefaction waves, which are continuous self-similar solutions of the
hydrodynamic limit equations \cite{Kipn99}. Various properties of the
fundamental excitations like stability, speed and morphology are determined by
a scalar or vector function which relates steady macroscopic currents to
average particle densities, the so-called current density relation. The
topology of the current-density function (or surfaces, in case of several
species of particles) such as the number of extrema and saddle points
determines qualitative features of the large scale dynamics and in particular
the number and character of the different stationary phases and phase
transitions that one can observe in the underlying microscopic model
\cite{Popk99,PopkovCambridge}. In this way microscopic details of local
particle interactions are largely irrelevant as long as they produce a certain
type of a current density relation.

In this work we identify a new large-scale excitation, reminiscent of a shock
wave, but which should be unstable according to usual shock stability
criteria. Focussing on models with two particle species we relate its
appearance to a special property of the current-density relation, the presence
of an isolated umbilic point. We shall call therefore this new excitation an
umbilic shock, or a U-shock. The aim of this article is to describe
microscopic and macroscopic properties of the U-shock, and to investigate
conditions for its appearance and stability. We find that such an excitation
is not at all exotic and can generally be observed in bidirectional models
with left-right symmetry in the hopping rates, provided that there is a
sufficiently strong repulsive interaction.

The plan of the paper is the following: In Sec.
\ref{sec::The model and a microscopic U-shock} we introduce our model and
describe the U-shock microscopically, highlighting its difference from a usual
shock. In Sec. \ref{sec::Umbilic point} we discuss macroscopic current density
relations with an umbilic point which makes the existence of the U-shock
possible. In Sec.\ref{sec::Umbilic point in bidirectional models} we prove
that bidirectional models with left-right symmetry in the hopping rates all
necessarily have an umbilic point. We finish with conclusions and
perspectives. The Appendices contain necessary technical details.

\section{The model and a microscopic U-shock}

\label{sec::The model and a microscopic U-shock}

Our model describes particles with repulsive hard-core interaction which hop
unidirectionally along two chains of $L$ sites: One chain for right-hopping
particles and another chain for left-hopping particles. At each instant of
time the system is fully described by occupation numbers $n_{k} \in\{0,1\} $
(for the right movers) and $m_{k} \in\{0,1\}$ (for the left-movers). A
right-moving particle at site $k$ can hop to its neighbouring site $k+1$
provided it is empty, with a rate that depends on the occupancies at sites
$k,k+1$ on the adjacent chain, see Fig.1. E.g. a particle hops with rate
$\beta$ if the adjacent sites are both occupied, etc. For clarity of
presentation and analytic simplification we shall keep only one rate
$\gamma=\mathrm{e}^{\nu}$ different from others, setting all remaining rates
to $1$,%
\begin{equation}
\alpha=\beta=\varepsilon=1, \quad\gamma= \mathrm{e}^{\nu} \label{rates}%
\end{equation}
Then the parameter
\begin{equation}
Q=\gamma-1 \label{InteractionQ}%
\end{equation}
which ranges from $\ -1$ to $\infty$, measures the interaction strength
between the left- and right-moving species. For $Q=0$ the model reduces to two
independently running totally asymmetric exclusion processes. The reason for
the given choice of rates is a simplification that it offers: The
current-density relation can be found analytically as explained in Sec.
\ref{sec::Umbilic point} and can therefore be analyzed in detail. For
monitoring the microscopic position of shocks on the lattice we also introduce
a second class particle (SCP) \cite{SCP}. This is a phantom particle which is
located at some site k (not disturbing real particles) and hops to the right
$k \rightarrow k+1$ if it finds a configuration $n_{k+1}=1,m_{k+1}=0$ on its
right, or it hops to the left $k \rightarrow k-1$. It is well-known that such
dynamical rules favor positioning of the SCP at the middle of a local density
gradient that corresponds to a shock on macroscopic scale.

\begin{figure}[ptb]
\centerline{\scalebox{0.6}{\includegraphics{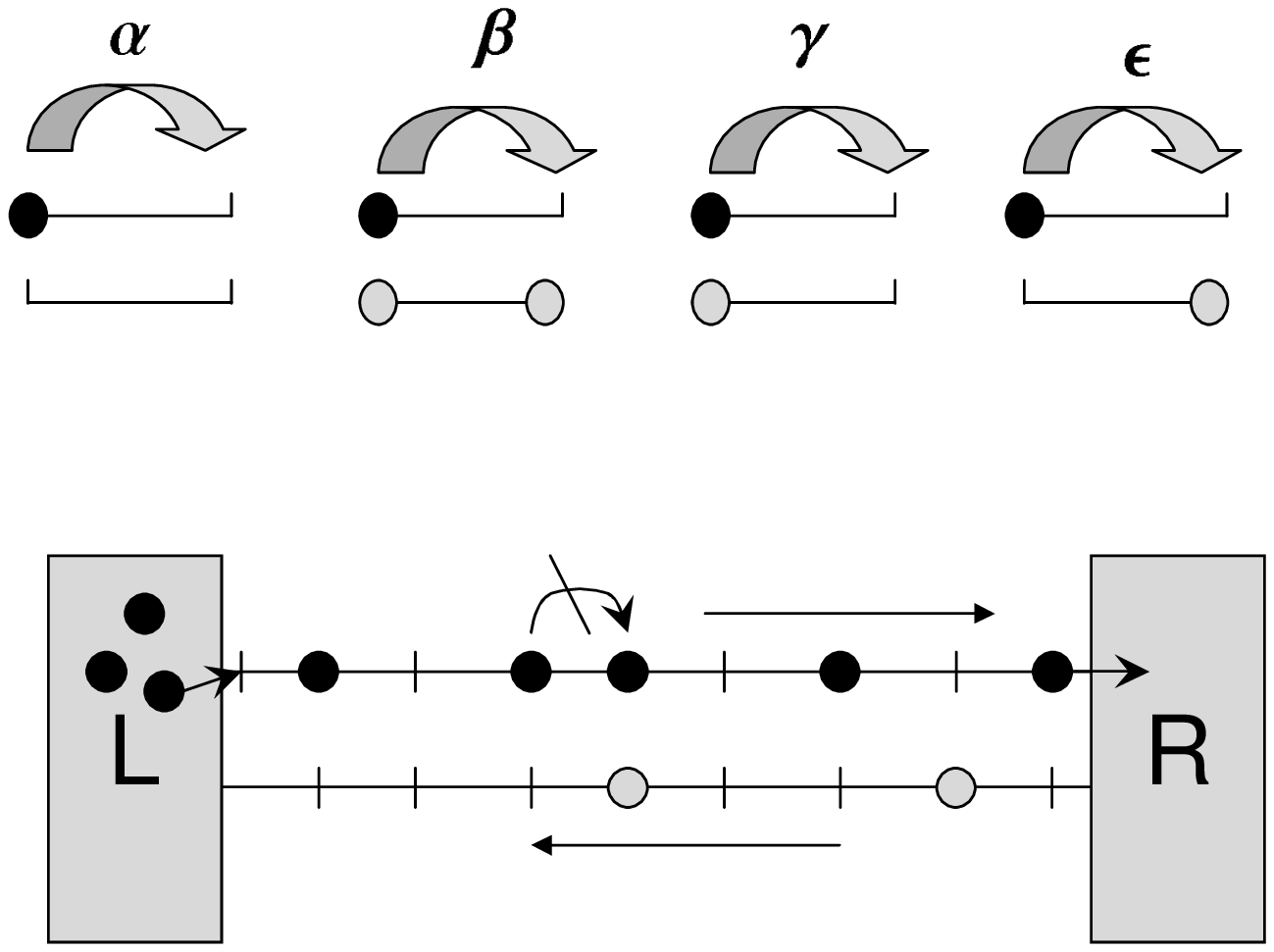}}
}\caption{Bidirectional two-chain model. For solvability,the rates must
satisfy $\alpha=\beta=\varepsilon=1$, $\gamma= \mathrm{e}^{\nu}$ where $\nu$
is the interchain interaction constant \cite{GunterSlava_StatPhys03}. Coupling
to boundary reservoirs is indicated by boxes marked L (the left reservoir) and
R (the right reservoir).}%
\label{Fig_BidirectionalModel}%
\end{figure}

The bulk dynamics is complemented with boundary conditions: We consider open
boundaries where at the left end of the chain a right mover can enter the
chain and it can leave it at the right end. Left movers are hopping to the
left with the same dynamic rules. Note that in general we do not require
complete left-right symmetry so the entrance and exit rates for different
species can be different. We choose a maximal feeding regime where we put a
particle on the entrance site once it becomes empty and take it out from the
exit site once it reaches it. The unidirectional hopping along with the open
boundaries ensure that a non-zero steady state current is maintained.

Our results do not depend qualitatively on how exactly the maximal feeding
regime is realized. For our dynamical Monte-Carlo simulation we choose the
following random sequential update procedure. For a chain of length $L$, i.e.
a system of $2L$ sites (numbered $i=1,2,...L$ for right movers and
$i=L+1,L+2,...2L$ for left movers) one Monte-Carlo step consists of $2L+2$
uniform drawings of a random number $s$ in the range $0\leq s\leq2L+1$. If $s$
falls within a segment $[0,L]$, the configuration of right movers is updated.
If $s=0$, and the left boundary site $i=1$ is empty, we fill it with a
particle (free entrance). If $s=L$ and the respective site contains a
particle, we remove it (free exit). For intermediate $0<s<L$, if site $s$
contains a particle, a hopping is performed on the right neighbouring site
with given rates (\ref{rates}), provided it was empty. The update of the left
movers is done analogously. We start from an empty lattice and after a
transient time we measure site occupancies $n_{k},m_{k}$, and take averages
over many Monte Carlo steps and many histories. Typically we choose a system
size up to $L=1000$ sites in each chain. The transient time for $L=1000$ is
$10^{6}$ Monte Carlo steps, and averaging over up to $10$ histories is done.
We perform the measurements for different values of the interaction parameter
$Q$. Note that due to the hardcore exclusion, the average densities of the
right and left-moving particles may only take values between $0$ and $1$.

\begin{figure}[ptbh]
\begin{center}
\subfigure[\label{Fig_Profiles::a}]
{\includegraphics[width=0.8\textwidth]{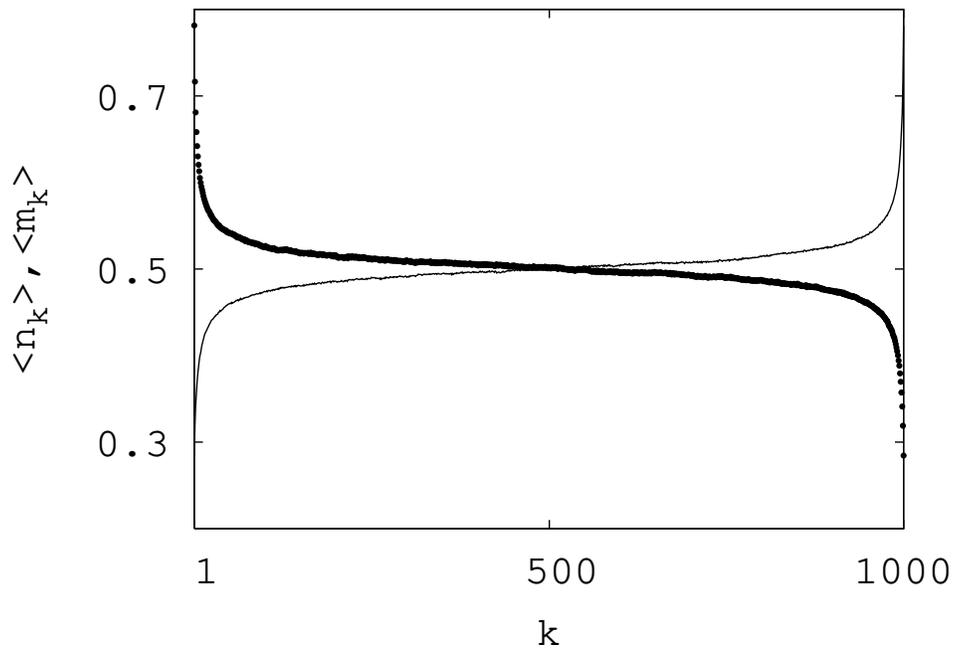}} \qquad
\subfigure[\label{Fig_Profiles::b}]{\includegraphics[width=0.8\textwidth]{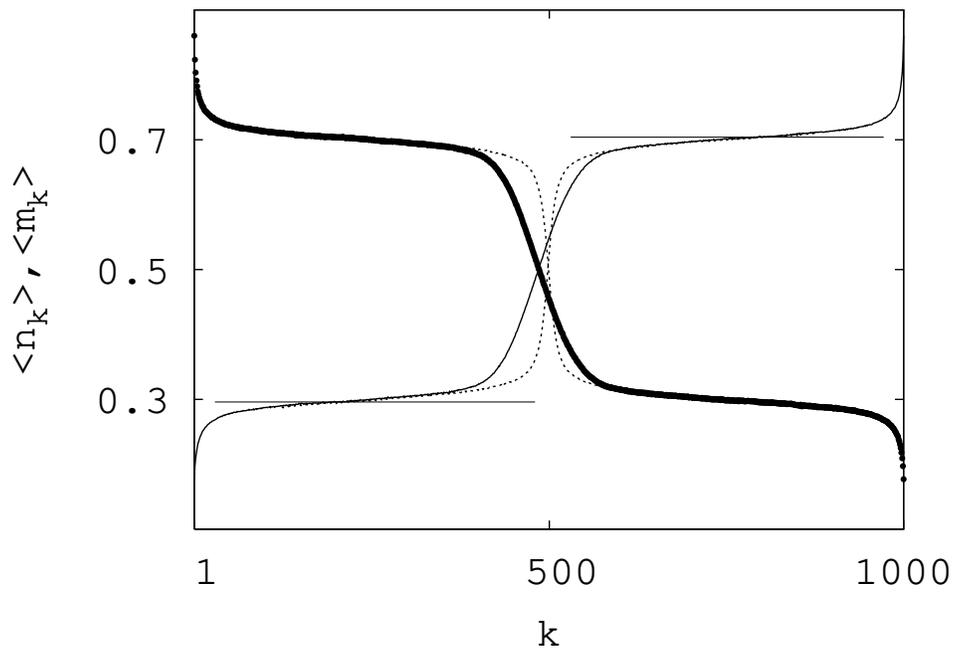}}
\end{center}
\caption{Average density profiles for right movers (thick lines) and left
movers (thin lines), above the phase transition $Q=-0.4$ (Panel(a)) and below
the phase transition $Q=-0.9$ (Panel(b)). The dotted curve on Panel(b) shows
the U-shock viewed from a second class particle. The U-shock profile, seen
from the SCP position, does not depend on the system size (data not shown).
Straight line pieces, showing bulk densities predicted from Eqs.
(\ref{UstatRHO1}),(\ref{UstatRHO2}), serve as a guide for an eye.}%
\label{Fig_Profiles}%
\end{figure}

The maximal feeding regime usually leads to the largest particle current since
we facilitate maximally the entrance and exit of particles at the boundary. In
the absence of interaction ($Q=0$) such boundary conditions lead to a state
with average particle densities $1/2$ a state with maximal possible particle
current \cite{Schu93,Derr93}. In the presence of interaction $Q$ the
stationary density profile does not undergo qualitative changes for a vast
interaction range $-0.75<Q<\infty$, see Fig.\ref{Fig_Profiles::a}. However,
for values of $Q<-0.75$ one observes something very unusual and different. The
bulk density profile becomes inhomogeneous and consists of two plateaux with
an interface in the middle. The profiles of the two species are left-right
symmetric but in each plateau the densities $\rho_{1},\rho_{2}$ of the left-
and right-movers are different, see Fig.\ref{Fig_Profiles::b}. As the
interaction becomes stronger, the difference $\rho_{1}-\rho_{2}$ grows and
reaches the maximum $\rho_{1}-\rho_{2}=0.5$ for the extreme case $Q=-1$. Note
that the asymmetry of the profile is not a result of a spontaneous symmetry
breaking since the profiles are left-right symmetric and the stationary
currents of both species remain equal.

What is the nature of the observed state? The maximal feeding regime in a
bulk-driven particle system usually produces steady states that are controlled
by rarefaction waves which (in an infinite system) are self-similar solutions
of the type $\rho(x,t)=\rho(\frac{x-x_{0}}{t})$, see also Appendix A. However,
the interface in the middle cannot be a rarefaction wave because it does not
change with time. With an increase of the system size, the interface gets
wider. However the widening is due to a fluctuation of a position of the
interface, as an analysis using second class particle shows, see
Fig.~(\ref{Fig_Profiles::b}). This is a property typical of a shock.

\begin{figure}[ptb]
\begin{center}
\includegraphics[
height=4.0594in, width=5.7943in ]{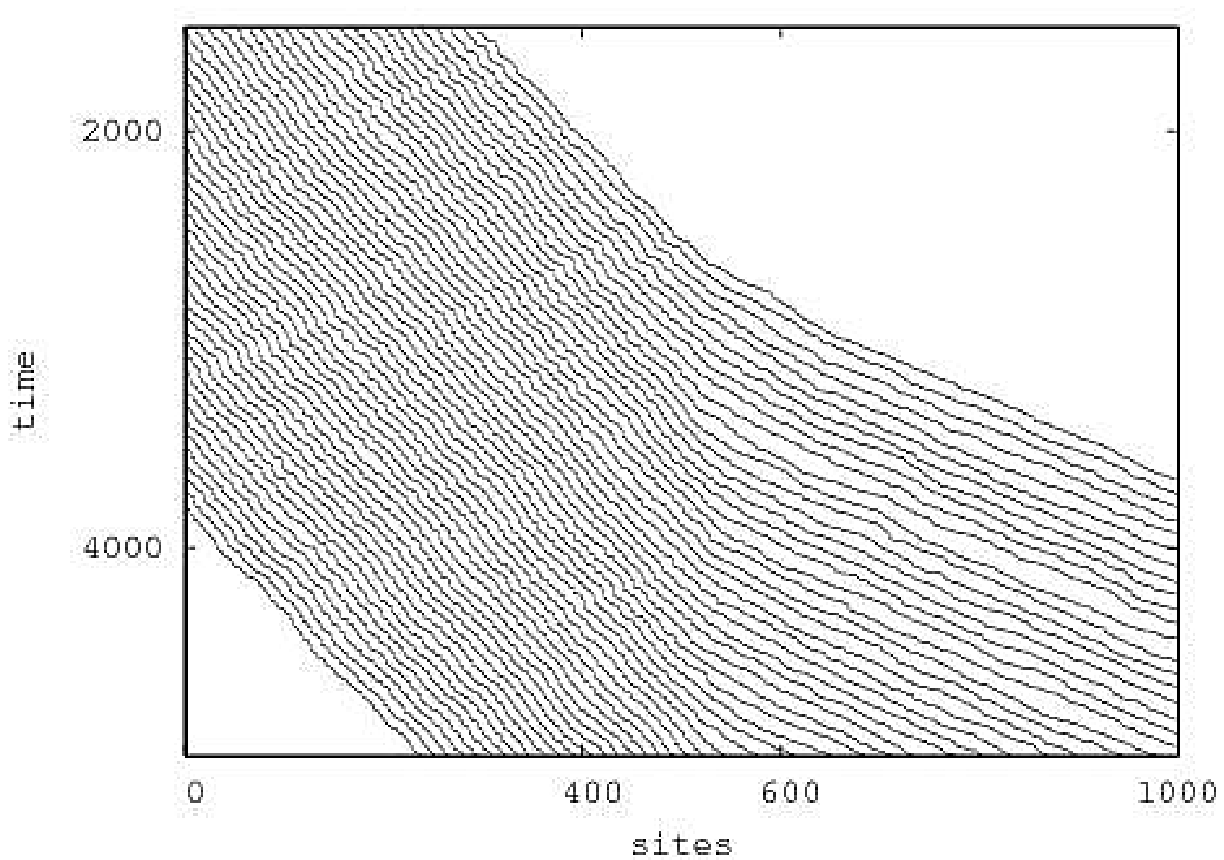}
\end{center}
\caption{Space-time trajectories of the right movers across a U-shock. Every
$10$-th trajectory is shown. A system of $1000$ sites was equilibrated for
$2\ast10^{6}$ Monte Carlo Steps before the trajectories were recorded. The
parameters are: $Q=-0.9$. }%
\label{Fig_TrajectoriesQm09}%
\end{figure}

However, the interface we observe is not a usual shock, either. In order to
see this, it is instructive to look at individual particle trajectories across
the interface, see Fig.~(\ref{Fig_TrajectoriesQm09}). Unlike the trajectories
running across a shock, the particles are moving slowly (in environment of
large density) across the left side of the lattice, and then accelerate after
crossing the inhomogeneity to the right-hand side of the lattice. Moreover,
according to usual shock stability conditions, using characteristic velocities
(see details in Sec.~\ref{sec::Umbilic point}), the interface shown on
Fig.~(\ref{Fig_Profiles}) should be unstable. As we shall argue below, the
reason for the stability and existence of the new state is an isolated umbilic
point in the current-density relation.

\section{Umbilic point in a current-density relation}

\label{sec::Umbilic point}

The model (\ref{rates}) that we consider has the remarkable property that the
stationary distribution is a product measure \cite{GunterSlava_StatPhys03}.
The steady state probabilities of any configuration $C$ in a periodic system
are given by
\begin{equation}
P_{C}=Z^{-1}%
{\displaystyle\prod\limits_{k=1}^{L}}
e^{-\nu n_{k}m_{k}},\label{ProductMeasure}%
\end{equation}
where $n_{k},m_{k}$ are particle occupation number on site $k$ on chains $1$
and $2$. With (\ref{ProductMeasure}) the stationary currents $j_{1}$ and
$j_{2}$ of the right- and left-moving species can be calculated exactly as
\begin{align}
j_{1}(u,v) &  =u(1-u)+Q\Omega_{11}(u,v)\Omega_{00}(u,v)\label{Ju}\\
j_{2}(u,v) &  =-v(1-v)-Q\Omega_{11}(u,v)\Omega_{00}(u,v)\nonumber
\end{align}
where $u$ and $v$ are the average densities of the right and of the left
movers. The quantities $\Omega_{11}$ and $\Omega_{00}$ are are stationary
probabilities to have two adjacent particles and two adjacent holes, given by
\begin{align}
\Omega_{11} &  =\frac{(u+v-1)Q-1+\sqrt{\left(  (u+v-1)Q-1\right)  ^{2}+4Quv}%
}{2Q}\label{Omega11}\\
\Omega_{00} &  =1-u-v+\Omega_{11}.\nonumber
\end{align}

From the stationary currents we construct the flux Jacobian $(Dj)$%
\begin{equation}
(Dj)=%
\begin{pmatrix}
\frac{\partial j_{1}}{\partial u} & \frac{\partial j_{1}}{\partial v}\\
\frac{\partial j_{2}}{\partial u} & \frac{\partial j_{2}}{\partial v}%
\end{pmatrix}
. \label{Jacobian}%
\end{equation}
Its two eigenvalues $c_{1,2}(u,v)$ play a fundamental role as characteristic
speeds of the system of conservation laws
\begin{align}
\partial_{t} u + \partial_{x} j_{1}(u,v)  &  = 0\\
\partial_{t} v + \partial_{x} j_{2}(u,v)  &  = 0\nonumber
\end{align}
which describes the coarse-grained dynamics on macroscopic scale.
Microscopically the characteristic speeds are the velocities of the localized
perturbations of a stationary homogeneous background with densities $u,v$
\cite{GunterSlava_StatPhys03}. As such, they determine stability of shocks and
rarefaction waves in the system.

A commonly made assumption about the flux functions $j_{1},j_{2}$, called
strict hyperbolicity, reads: the characteristic speeds are different
$c_{1}(u,v)$ $\neq c_{2}(u,v)$ for all $u,v$. Strictly hyperbolic systems have
only two types of fundamental solutions: shocks and rarefaction waves
\cite{Lax2006}. As argued in the previous section, the $U$-shock is neither a
usual shock nor a rarefaction wave, so it cannot not be stable in a strictly
hyperbolic system.

Indeed, our system is not a strictly hyperbolic one, but it has a so-called
umbilic point which is defined as a point in the $u-v$ density plane where
where the two characteristic velocities coincide. It is straightforwardly
verified from the analytic expressions for the currents that for our system
this is the case at $u^{\ast}=v^{\ast}=1/2$, where the two characteristic
speeds $c_{1,2}^{\ast}$ are equal and zero for all values of $Q$.

\begin{figure}[ptbh]
\begin{center}
\subfigure[\label{Fig_Qneg}]
{\includegraphics[width=0.45\textwidth]{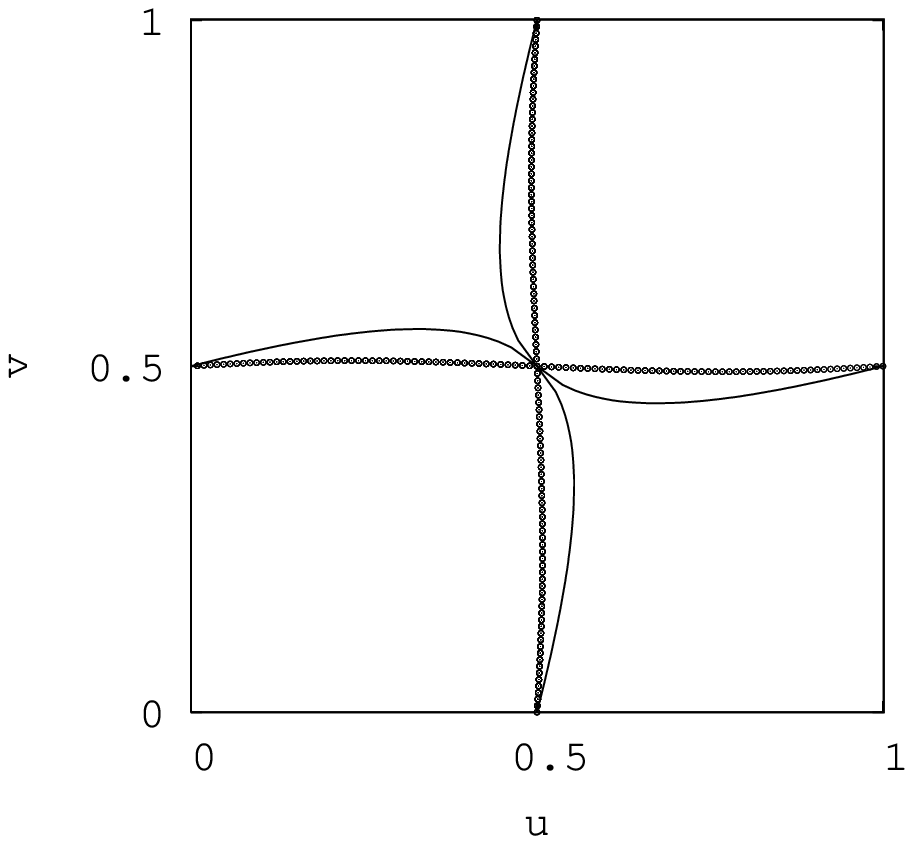}} \qquad
\subfigure[\label{Fig_Qcrit}]
{\includegraphics[width=0.45\textwidth]{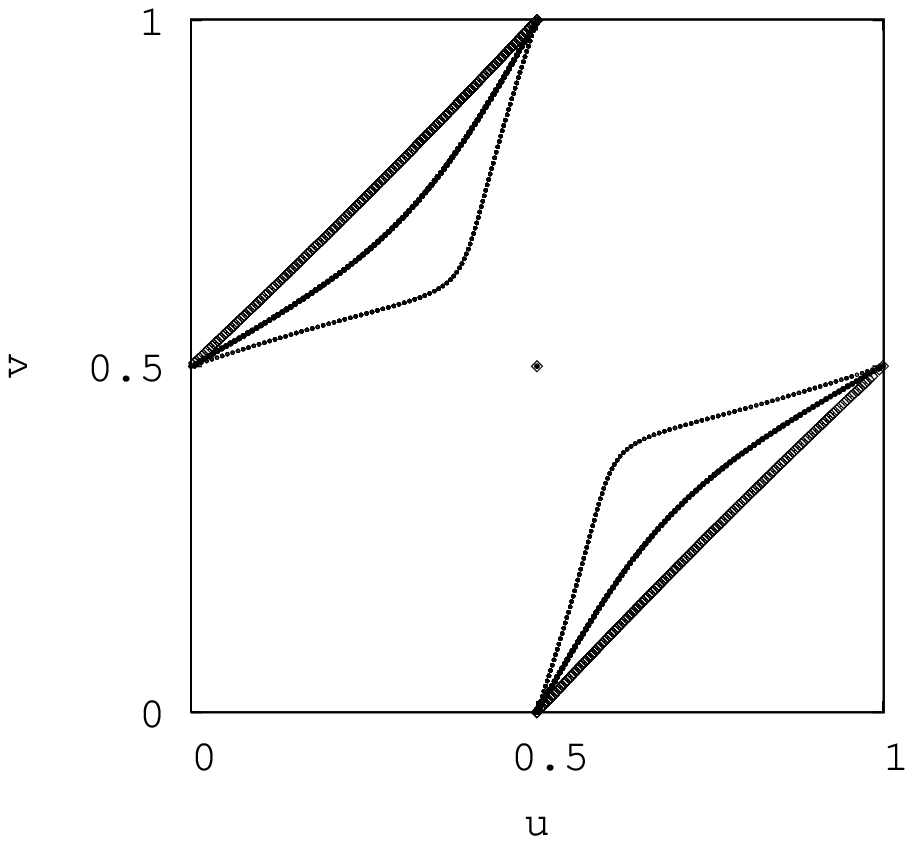}}
\end{center}
\caption{ Location of the curves in $u-v$ plane where at least one
characteristic velocity vanishes $c_{i}(u,v)=0$, for small negative
$Q=-0.5,-0.75$( \textbf{Panel (a)}, thick and thin lines respectively) and
large negative $Q=-0.8,-0.9,-0.99$( \textbf{Panel (b)}, thin, medium and thick
lines respectively). The point $u=v=1/2$ is an umbilical point where
$c_{1}=c_{2}=0$ for any value of $Q$. For $Q<-0.75$, the umbilical point
becomes an isolated point. }%
\label{FigOmegaQ}%
\end{figure}

For a full discussion of the current-density relation (\ref{Ju}) and the
associated flux Jacobian (\ref{Jacobian}) we note that the points where one
characteristic speed vanishes generically correspond to a family of
rarefaction waves \cite{Lax2006,Lax73}, see also Appendix A. Looking at the
location of the points where at least one characteristic speed vanishes, we
find two different topologies, depending on $Q$. For the interaction range
$-3/4<Q<\infty$ the umbilic point $(u^{\ast},v^{\ast})$ is a crossing point of
the curves $c_{1}(u,v)=0$ and $c_{2}(u,v)=0$. For $-1\leq Q<-3/4$, the umbilic
point $(u^{\ast},v^{\ast})$ is an isolated point, and the curves
$c_{k}(u,v)=0$ do not cross, see Fig.~(\ref{FigOmegaQ}).

\begin{figure}[h]
\centerline{\scalebox{0.6}{\includegraphics{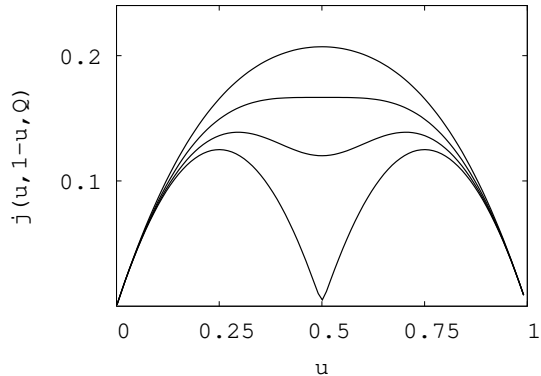}}
}\caption{Bidirectional two-chain model. Cuts of the stationary current
surface along $v=1-u$ line: $j_{1}(u,1-u,Q)$ at different
$Q=-0.5,-0.75,-0.9,-0.9999$ (curves up to down). The cuts of the stationary
current surface along the perpendicular direction $u=v$ remain convex for all
values of $Q$ (data not shown).}%
\label{Fig_CurrUV}%
\end{figure}

The appearance of the \textit{isolated} umbilic point is a consequence of a
change of topology of the current surfaces $j_{k}(u,v,Q)$ from a convex to a
saddle point shape at a critical value of $Q_{cr}=-3/4$. To understand from a
microscopic perspective how this happens consider a cut of the current surface
along the line $v=1-u$ for small $\gamma=Q+1\ll1$. At half-filling $u=v=1/2$
the system comes into a configuration where all adjacent sites are either both
occupied or both empty and gets essentially stuck: all hoppings from this
configuration are suppressed by a small hopping rate $\gamma\ll1$. This strong
repulsive interaction between the particles on the two lanes explains the
occurrence of a double maximum in the curve $j(u,1-u,Q)$ for $Q<Q_{cr}$ as
shown on Fig.\ref{Fig_CurrUV}. The appearance of the double maximum then gives
rise to the isolated umbilic point. The positions and amplitudes of the
extrema of $g(u,Q)=j(u,1-u,Q)$ are readily found from (\ref{Ju}),
(\ref{Omega11}): for $Q>Q_{cr}$ the $j(u,1-u,Q)$ curve has one maximum at
$u_{0}^{\ast}=1/2$, while for $-1\leq Q<Q_{cr}$ it has three extrema at
positions $u_{0}^{\ast}=1/2$ and $u_{1,2}^{\ast}=\frac{1}{2}\pm\sqrt
{3Q^{-1}+4}/4$. The respective currents are $j_{0}^{\ast}=\sqrt{Q+1}%
/(2\sqrt{Q+1}+2)$ and $j_{1}^{\ast}=j_{2}^{\ast}=1/(8|Q|)$.

Now we are in a position to analyze the U-shock solution. We can identify the
average bulk densities of the left plateau $u_{-},v_{-}$ of the U-shock (see
Fig.2) with $u_{1}^{\ast},u_{2}^{\ast}$ respectively, for the following
reasons: (A) We expect the U-shock plateaux, piecewise, to be governed by a
rarefaction wave, meaning that at least one of characteristic speeds must
vanish, $c_{i}(u_{-},v_{-})=0$. (B) We expect the stationary currents
amplitudes for right and left-moving species to be equal, due to left-right
symmetry, i.e. $j_{1}(u_{-},v_{-})=-j_{2}(u_{-},v_{-})$. We readily find,
using (\ref{Ju}),(\ref{Omega11}), three pairs of solution satisfying (A)and
(B): (i)$(u_{-},v_{-})=(1/2,1/2);$ (ii) $(u_{-},v_{-})=(u_{1}^{\ast}%
,u_{2}^{\ast})$ and (iii) $(u_{-},v_{-})=(u_{2}^{\ast},u_{1}^{\ast})$.
Comparing with the U-shock, we find $(u_{-},v_{-})=(u_{1}^{\ast},u_{2}^{\ast
})$ to be the relevant solution. Indeed the solution (iii) is not compatible
with our boundary conditions, while the solution (i) would result in a
reduction of the particle current and is dynamically unstable.

For the right plateau of the U-shock, we find analogously $(u_{+}%
,\,v_{+})=(u_{2}^{\ast},\,u_{1}^{\ast})$. Apparently, the first solution (i)
is unstable for $Q<Q_{cr}$. So, we have for $Q<Q_{cr}$
\begin{align}
u_{-}(Q)  &  =v_{+}(Q)=\frac{1}{2}+\frac{\sqrt{3Q^{-1}+4}}{4}\label{UstatRHO1}%
\\
v_{-}(Q)  &  =u_{+}(Q)=\frac{1}{2}-\frac{\sqrt{3Q^{-1}+4}}{4}\text{.}
\label{UstatRHO2}%
\end{align}
The stationary currents for the U-shock are then given by the $j_{k}%
(u_{1}^{\ast},u_{2}^{\ast},Q)$,
\begin{equation}
j_{stat}^{U}=\frac{1}{8|Q|}\text{ for }Q<-\frac{3}{4}\text{.}
\label{CurrSTATsmallQ}%
\end{equation}
Above the critical point, the bulk densities are $u=v=1/2$, and the stationary
currents are given by $j_{k}(1/2,1/2,Q)$,
\begin{equation}
j_{stat}=\frac{\sqrt{Q+1}}{2\sqrt{Q+1}+2}\text{ for }Q\geq-\frac{3}{4}.
\label{CurrSTATlargeQ}%
\end{equation}

\begin{figure}[h]
\centerline{\scalebox{0.6}{\includegraphics{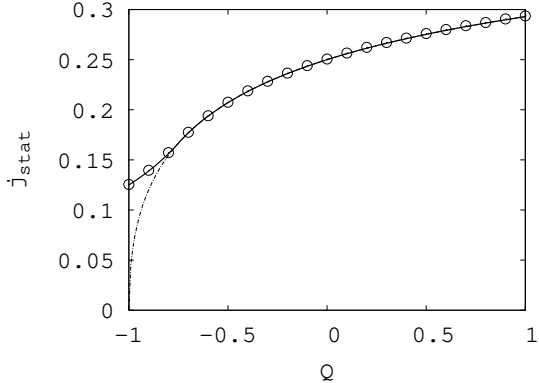}}
}\caption{Stationary currents of the bidirectional model at maximal feeding as
function of $Q$. Circles show Monte-Carlo results for a system of size 600,
while solid curve is the theoretical prediction (\ref{CurrSTATsmallQ}%
),(\ref{CurrSTATlargeQ}). Below the critical $Q$, the branch
(\ref{CurrSTATlargeQ}), indicated by the broken line, becomes unstable.
(\ref{CurrSTATlargeQ}). }%
\label{Fig_QcurrQSTAT}%
\end{figure}

Our analytical predictions are well borne by the MC simulations, see
Fig.\ref{Fig_QcurrQSTAT}. Note that by establishing an inhomogeneous state
(the U-shock) below $Q_{cr}$ the system optimizes its current, which would be
strongly suppressed for any symmetric bulk homogeneous state. This can be
viewed as a generalization of the phenomenon of a current maximization at
maximal feeding \cite{Krug91} to a system with two species.

We can study the phase transition in our model at $Q=Q_{cr}$. Let us choose
the difference $u_{-}(Q)-v_{-}(Q)=\Delta$ between the bulk densities of the
right and left movers to be our order parameter. We have $\Delta=0$ for
$Q>Q_{cr}$ and $\Delta=\frac{1}{2}\sqrt{3Q^{-1}+4}$ for $Q<Q_{cr}$. Near the
transition, $\left.  \Delta\right\vert _{Q_{cr}-\delta q}=\sqrt{\delta q/3}$,
so we have a square root singularity, similar to that arising in the Landau
theory of continuous phase transitions generated by a change of a free energy
potential from a single to a double minimum topology. On the other hand, the
stationary current is continuous across the transition point together with its
first derivative,
\[
\left.  j_{stat}\right\vert _{Q_{cr}-\delta q}-\left.  j_{stat}\right\vert
_{Q_{cr}}=O(\delta q^{2}),\text{ for }\delta q\ll1\text{.}%
\]

Finally, we comment on robustness of the U-shock. The U-shock turns out to be
very robust with respect to a choice of the boundary conditions (BC). We
observe the U-shock for a wide choice of BC, which in particular do not need
to be left-right symmetric. In order to formulate the conditions for its
appearance more precisely, we need to parametrize the BC in terms of boundary
reservoirs. Such a parametrization involves further technical details
\cite{reflections_ABO} and is out of the scope of the present paper.

The U-shock is also stable w.r. t. a change of the model parameters, e.g. the
bulk hopping parameters, as long as they remain left-right symmetric. In
particular, the particle-hole symmetry of the hopping rates (\ref{rates}),
which makes the particle current invariant w.r.t. $u,v\rightarrow1-u,1-v$
interchange (see Fig.\ref{Fig_CurrUV}) can be relaxed without causing
qualitative changes to a U-shock.

\section{Umbilic point in bidirectional models}

\label{sec::Umbilic point in bidirectional models}

With the system we have studied, we were lucky enough to find the stationary
currents analytically and establish the existence of the umbilic point. How
exotic is this point? Moreover, is it possible to predict, from the
microscopic transition rates, if the system will have such a point? In this
section we prove that an umbilic point with $c_{1}=c_{2}=0$ is not at all
exotic and is present necessarily in bidirectional models with left-right symmetry.

Let us consider Markov processes involving two driven particle species which
are biased in opposite directions, their bulk dynamics (but not necessarily
their boundary dynamics) being invariant under the left-right interchange. Let
us denote the average particle densities of the two species as $u,\,v$ and the
respective currents as $j_{1}(u,v),\,j_{2}(u,v)$. From the left-right symmetry
we have $j_{2}(u,v)=-j_{1}(v,u)$. Let us consider the line of equal densities
$u=v$. Along this line the flux Jacobian (\ref{Jacobian}) takes the form
\[
Dj=%
\begin{pmatrix}
a(u) & b(u)\\
-b(u) & -a(u)
\end{pmatrix}
\]
with the respective eigenvalues (characteristic speeds)
\begin{equation}
c_{1}(u=v),c_{2}(u=v)=\mp\sqrt{a^{2}-b^{2}}. \label{c1=-c2}%
\end{equation}
Let us assume, in addition, that we have a restriction on a number of
particles which can occupy a single lattice site, i.e. that the maximally
allowed density of the each species is limited to the same value $\max u=\max
v=\rho_{\max}$. In many applications such a restriction is a consequence of
the hard core exclusion interaction.

Now, let us move along the line $u=v$ from $u=0$ to the maximally allowed
value $u_{\max}$ and assume without loss of generality that our process
evolves on two parallel chains, $u_{\max}=v_{\max}=\rho_{\max}$. Let us call
the lane with right movers the lane $A$, and the lane with left movers the
lane $B$.

Guided by the physical meaning of the characteristic speeds as the velocities
of localized perturbations of a homogeneous state
\cite{GunterSlava_StatPhys03}, we deduce that a stationary state with small
density of right movers on one lane and small density of left movers on
another lane, attainable for $u=v\rightarrow0$, has characteristic speeds of
opposite signs in accordance with (\ref{c1=-c2}). In this limit the left and
right moving species are practically uncoupled and we can attribute the
positive characteristic speed to right-moving particles $c_{A}(u=v\rightarrow
0)>0$ and the negative characteristic speed to the left-moving particles
$c_{B}(u=v\rightarrow0)<0$. At the other end of the line $u=v\rightarrow
\rho_{\max}$ we have vanishing density of left moving holes in the dense
background of the right moving particles on lane A, and similarly for the
right moving holes on the lane B. Repeating the arguments for the
characteristic speeds, we have $c_{A}(u\rightarrow u_{\max})<0$ $\ $\ and
$c_{B}(u\rightarrow u_{\max})=-c_{A}(u\rightarrow u_{\max})$. By continuity we
deduce that there exists a point at which the characteristic speed $c_{A}$
changes sign $c_{A}(u^{\ast}=v^{\ast})=0$. Moreover, due to (\ref{c1=-c2}),
the other characteristic speed at this point, also vanishes $c_{B}(u^{\ast
}=v^{\ast})=0$. The point $u^{\ast}=v^{\ast}$ where both characteristic speeds
vanish is an umbilic point.

This argument establishes the existence of at least one umbilic point with
$c_{1}=c_{2}=0$ for left-right symmetric bulk dynamics. Notice that the
boundary conditions do not enter the argument. Hence one may impose boundary
conditions that violate the left-right symmetry with destroying the umbilic point.


\section{Conclusions}


We have considered an open bidirectional two-component driven diffusive system
with left-right symmetry for the bulk dynamics in the maximal flow regime. We
have discovered and described in detail a bulk inhomogeneous solution of a
novel type, denoted U-shock. This solution has many properties of a usual
shock -- in particular its microscopic sharpness -- but does not satisfy the
usual criteria for the stability of shocks. We have computed the critical
value of the (repulsive) interaction above which the U-shock exists. We have
shown that the existence of the U-shock is due to the intrinsic presence of an
isolated umbilic point with vanishing characteristic velocities. No
fine-tuning of the interaction strength is required above the critical value.

The U-shock turns out to be robust also with respect to changes in the
boundary parameters, even if they violate the (necessary) left-right symmetry
of the bulk dynamics. In an open system it is only necessary maintain a
stationary a maximal flow regime. More generally, we proved the existence of
an umbilic point with vanishing characteristic velocities in any bidirectional
model with left-right symmetry of the bulk hopping rates. The presence of
umbilic points alters crucially the dynamics of the system in the hydrodynamic
limit (\ref{PDE}), giving rise to a variety of unusual solutions, called
undercompressed and overcompressed shocks \cite{ChenKan95}. The U-shock is one
of such solutions. The necessary condition for an U-shock is a sufficiently
strong repulsive interaction, resulting in a saddle point in the
current-density surfaces.

Bidirectional models are being widely studied in the literature, in
particular, in connection with the intriguing phenomenon of spontaneous
symmetry breaking (SSB) \cite{bridge}-\cite{bridgeJiang}. However, we find
that in most known cases the current-density relations are convex surfaces. It
would be interesting to study SSB in presence of an isolated umbilic point.
Another interesting problem is to explore the full phase diagram of the open
system and to find out which role the U-shock might play in boundary driven
phase transitions.

\section*{Acknowledgements}

V.P. thanks the IZKS and the University of Bonn for hospitality and
acknowledges a partial support by the Alexander von Humboldt foundation.

\appendix

\section{Rarefaction-wave controlled stationary states}

\label{Appendix::Rarefaction-wave controlled stationary states}

The generical importance of the points where at least one of characteristic
speeds $c_{i}$ vanishes can be demonstrated by the following argument. It is
well-known \cite{Lax2006,Lax73} that partial differential equations of the
type%
\begin{align}
{\frac{\partial\rho_{k}}{\partial t}}+{\frac{\partial j_{k}(\rho_{1},\rho
_{2},...\rho_{K})}{\partial x}}  &  =0\label{PDE}\\
{k}  &  {=1,2,.K},\nonumber
\end{align}
where $K$ is the number of species, admit two fundamental classes of
solutions: shock waves and rarefaction waves. A rarefaction wave is a
self-similar solution of (\ref{PDE}), depending only on the ratio
$\xi=(x-x_{0})/t$ where $x_{0}$ is the position of its center, and $t>0$. We
argue that in the long-time (stationary) limit $t\rightarrow\infty$ the
stationary bulk density $\mathbf{\rho}_{stat}$ generated by a rarefaction wave
has zero characteristic speed $c_{p}(\mathbf{\rho}_{stat})=0$. Here
$\mathbf{\rho}(x,t)$ is a vector the components of which are the density
profiles $\rho_{1}(x,t),\rho_{2}(x,t),...,\rho_{K}(x,t)$ of the respective
species. By $\mathbf{\rho}_{stat}$ we denote the vector with stationary bulk
densities $\{\rho_{1}^{stat},\rho_{2}^{stat},...,\rho_{K}^{stat}\}$.

We search for a solution of (\ref{PDE}) in the form $\mathbf{\rho}%
(x,t)=h(\xi)$, Substituting in (\ref{PDE}), we obtain
\begin{equation}
-\frac{\xi}{t}\frac{\partial h}{\partial\xi}+\frac{1}{t}(D\mathbf{j}%
)\frac{\partial h}{\partial\xi}=0
\end{equation}
where the matrix $(D\mathbf{j})$ is the Jacobian of the flux $(D\mathbf{j}%
)_{pq}=\partial j_{p}/\partial\rho_{q}$. The above equation can be rewritten
as
\begin{equation}
(D\mathbf{j})\frac{\partial h}{\partial\xi}=\xi\frac{\partial h}{\partial\xi}.
\end{equation}

In the limit $t\rightarrow\infty$, the scaled displacement $\xi=(x-x_{0}%
)/t\rightarrow0$ vanishes for any finite $x-x_{0}$, and the above equation
reduces to $\left.  (Dj)\right\vert _{t\rightarrow\infty}h^{\prime}=0$. Each
solution of this equation is an eigenvector of the flux Jacobian $D\mathbf{j}$
with zero eigenvalue. Consequently, the matrix $(Dj)_{t\rightarrow\infty
}=(D\mathbf{j})(\mathbf{\rho}_{stat})$ is a matrix with zero eigenvalue, i.e.
at the point $\mathbf{\rho}_{stat}$ at least one $c_{p}(\mathbf{\rho}%
_{stat})=0$. The respective rarefaction wave is called $p$-rarefaction wave
\cite{Lax2006,Lax73}. Of course, in order to guarantee the stability of the
above discussed rarefaction wave with respect to local perturbations at the
boundaries, the boundary conditions have to be chosen appropriately.


\end{document}